\documentclass{article}
\usepackage{LaThuileFPSpro}
\usepackage{graphicx}
\usepackage[super,sort&compress]{natbib}
\bibpunct{}{)}{}{}{}{}
\begin{document}

\newcommand{\pam}{\textsf{PAMELA} }
\newcommand{\antip}{$\overline{p}$}
\newcommand{\posit}{$e^+$}

\newcommand{\grl}{Geophysical Research Letters}
\newcommand{\prd}{Physical Review D}
\newcommand{\apj}{ApJ}
\newcommand{\apjl}{ApJ}
\newcommand{\aap}{A\&A}

\title{
  THE PAMELA COSMIC RAY SPACE  OBSERVATORY: DETECTOR,  OBJECTIVES and FIRST RESULTS
  }
\author{
Marco Casolino$^*$
 Daniel Bongue, Maria Pia De Pascale,
Nicola De Simone \\ Valeria Di Felice  Laura Marcelli   Mauro
Minori \\
Piergiorgio Picozza, Roberta Sparvoli\\
{\em INFN  and Physics Department
of University of Rome ``Tor Vergata" } \\
{\em  $^*$ Corresponding author: casolino@roma2.infn.it} \\
Guido Castellini\\
{\em IFAC, Florence, Italy } \\
 Oscar Adriani, Lorenzo Bonechi, Massimo Bongi \\  Sergio
Bottai
 Paolo Papini, Sergio Ricciarini\\  Piero Spillantini, Elena Taddei,
Elena Vannuccini\\
  {\em INFN,  and Physics Department of University of Florence } \\
Giancarlo Barbarino, Donatella Campana, Rita Carbone \\ Gianfranca
De Rosa,  Giuseppe Osteria \\
  {\em INFN,  and Physics Department of University of Naples ``Federico II" }
  \\
Mirko Boezio, Valter Bonvicini, Emiliano Mocchiutti, Andrea Vacchi \\ Gianluigi Zampa, Nicola Zampa\\
{\em INFN,  and Physics Department of
University of Trieste} \\
 Alessandro Bruno,
Francesco Saverio Cafagna\\
{\em INFN,  and Physics Department of University
of Bari} \\
Marco Ricci\\
{\em INFN, Laboratori Nazionali di Frascati,  Italy} \\
Petter Hofverberg, Mark Pearce, Per Carlson \\
{\em KTH,   Stockholm, Sweden} \\
Edward Bogomolov, S.Yu.~Krutkov, N.N.~Nikonov, G.I.Vasilyev   \\
 {\em Ioffe Physical Technical Institute, St. Petersburg, Russia} \\
 Wolfgang Menn, Manfred Simon\\
{\em Universit\"{a}t Siegen,    Germany } \\
 Arkady Galper, Lubov Grishantseva,  Sergey Koldashov, Alexey Leonov \\
   Vladimir V.~Mikhailov,   Sergey A.~Voronov,  Yuri T.~Yurkin, Valeri G.~Zverev \\
 {\em Moscow Engineering and Physics Institute,
 Moscow, Russia} \\
Galina A. Bazilevskaya, Alexander N. Kvashnin \\ Osman Maksumov, Yuri Stozhkov\\
{\em Lebedev Physical Institute, Moscow, Russia } \\
}
 \maketitle

\baselineskip=11.6pt

\begin{abstract}
  \pam\ is a satellite borne experiment designed to study with great accuracy
cosmic rays of galactic, solar, and trapped nature in a   wide
energy range (protons: 80 MeV-700 GeV, electrons 50 MeV-400 GeV).
Main objective is the study of the antimatter component:
antiprotons (80 MeV-190 GeV), positrons (50 MeV-270 GeV) and
search for antimatter with a precision of the order of $10^{-8}$).
The experiment, housed on board the  Russian Resurs-DK1 satellite,
was launched on June,
  $15$ 2006 in a $350\times 600~km$ orbit with an inclination of
70 degrees.  The detector  is composed of  a series of
scintillator  counters arranged at the extremities of a permanent
magnet spectrometer to provide charge, Time-of-Flight and rigidity
information.   Lepton/hadron identification is performed  by a
Silicon-Tungsten calorimeter and a Neutron detector placed at the
bottom of the device. An Anticounter system is used offline to
reject false triggers coming from the satellite. In self-trigger
mode the Calorimeter, the neutron detector and a shower tail
catcher are  capable of an independent measure of the lepton
component up to 2 TeV.  In this work we describe the experiment,
 its scientific objectives and the performance in its first two years of
 operation. Data  on protons of trapped, secondary and galactic
 nature - as well as measurements of the December  13 2006 Solar Particle Event - are provided.
 \end{abstract}
  \\
   Essentially a pre-antiparticle  measurement paper.
   \\
Invited talk in {\it La Thuile 2008},  Les Rencontres de Physique
de la Vall$\acute{e}$e d'Aoste. {\it Results and Perspectives in
Particle Physics}.
\\
24 February 2008 - 01 March 2008

\section{Introduction}

\label{sec:intro} The Wizard collaboration is a scientific program
devoted to the study of cosmic rays through balloon and
satellite-borne devices. Aims involve the precise determination of
the antiproton \cite{boe97} and positron \cite{boe00} spectrum,
search of antimatter, measurement of low energy trapped and solar
cosmic rays with the NINA-1 \cite{nina} and NINA-2 \cite{nina2}
satellite experiments. Other research on board Mir and
International Space Station has involved the measurement of the
radiation environment,  the nuclear abundances and the
investigation of the Light Flash \cite{nat} phenomenon with the
Sileye experiments \cite{sil2}\cite{sil3}. \pam\ is the largest
and most complex device built insofar by the collaboration, with
the broadest scientific goals.  In this work we describe the
detector, and its launch and commissioning phase. Scientific
objectives are presented together with the   report of the first
observations of protons of solar, trapped and galactic nature.

\section{Instrument Description}
\begin{figure}

\begin{center}
\includegraphics[width=.8\textwidth]{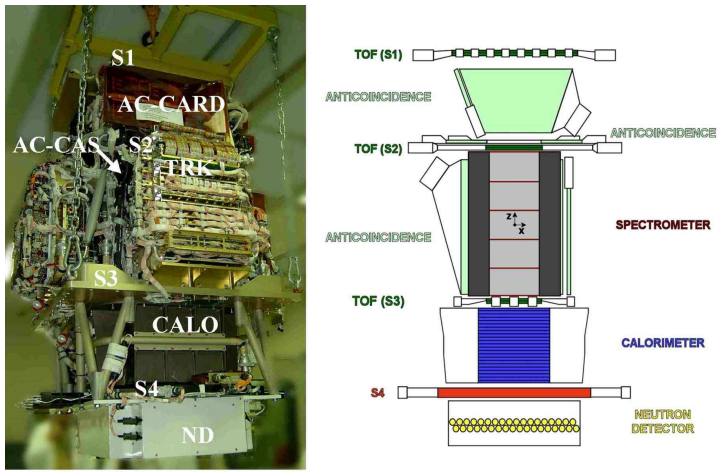}
 \caption{Left: Photo of
the  \pam\ detector during the final  integration phase in Tor
Vergata clean room facilities, Rome. It is possible to discern,
from top to bottom, the topmost scintillator system, S1,  the
electronic crates around the magnet spectrometer, the baseplate
(to which \pam\ is suspended by chains), the black structure
housing the Si-W calorimeter, S4 tail scintillator and the neutron
detector.  Right: scheme - approximately to scale with the picture
-  of  the detectors composing \pam. } \label{scheme2}
\end{center}
\end{figure}

\begin{figure}
\begin{center}
\includegraphics[width=0.8\textwidth]{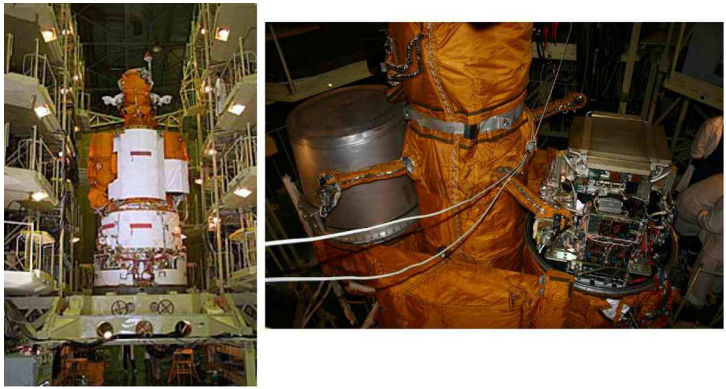}
  \caption{Left: Photo of
Resurs in the final integration phase in Baikonur. It is possible
to discern the   the optical sensor on top,  the two pressurized
containers on the sides, and the white heat cooling panel in the
forefront. Right: close up picture of the integration phase of
\pam\ in the pressurized container (right in picture).  }
\label{scheme1}
\end{center}
\end{figure}

In this section we  describe   the main characteristics of \pam\
detector; a more detailed description of the device and the data
handling  can be found in \cite{Pi07, cpu, yoda}. The device
(Figure \ref{scheme2}) is constituted by a number of highly
redundant detectors capable of identifying particles providing
charge, mass,  rigidity and beta    over a very wide energy range.
The instrument is built around a  permanent magnet  with a silicon
microstrip   tracker  with  a scintillator system to provide
trigger, charge and time of flight information. A silicon-tungsten
calorimeter is used to perform hadron/lepton separation. A shower
tail catcher and a neutron detector at the bottom of the apparatus
increase this separation. An anticounter system is used to reject
spurious events in the off-line phase. Around the detectors are
housed the readout electronics, the interfaces with the CPU and
all primary and secondary power supplies. All systems (power
supply, readout boards etc.)  are redundant with the exception of
the CPU which is more tolerant to failures. The system is enclosed
in a pressurized container (Figure \ref{scheme},\ref{scheme1})
located on one side of the Resurs-DK satellite. In a twin
pressurized  container is  housed the   Arina experiment, devoted
to the study of the low energy trapped electron and proton
component. Total weight of \pam\ is 470 kg;  power consumption is
355 W, geometrical factor is 21.5$cm^2 sr$.

\subsection{Resurs-DK1  Satellite}

\begin{figure}
\begin{center}
\includegraphics[width=0.8\textwidth]{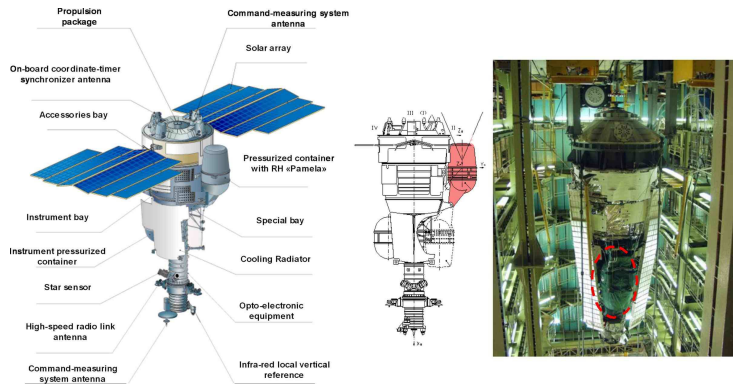}
  \caption{Left: Scheme of
the Resurs-DK1 satellite.   \pam\ is located in the pressurized
container on the right of the picture. In the center panel it is
possible to see the container in the launch position and in the
extended (cosmic ray acquisition) configuration. In the right
panel it is possible to see a picture of the satellite in the
assembly facility  in Samara. The picture   is rotated 180 degrees
to compare the photo with the scheme. The dashed circle  shows the
location of \pam\ pressurized container in the launch position.}
\label{scheme}
\end{center}
\end{figure}

The Resurs-DK1 satellite (Figure \ref{scheme}) is the evolution of
previous
 military reconnaissance satellites  flown   during in the  years 1980 - 1990. It
 was   developed by TsSKB Progress plant \cite{samara} in the city of Samara (Russia), in cooperation
  with NPP OPTEKS, OAO Krasnogorskiy Zavod, NIITP and NTsOMZ
 (Russia's Science Center
for Remote Sensing of Earth)\cite{ntsomz}.  The spacecraft is
three-axis stabilized, with axis orientation accuracy       0.2
arcmin and angular velocity stabilization accuracy of 0.005$^o$/s.
The spacecraft has a mass of about 6650 kg,   height of  7.4 $m$
and a solar array span of  about 14 m.   It is   designed to
provide imagery of the Earth surface for   civilian use and  is
the first Russian non-military satellite with   resolution
capability of $\simeq 0.8 $ m      in composite color
mode\footnote{Observations are performed in three bands ( 0.50 -
0.60$\mu m$ , 0.60 - 0.70$\mu m$, 0.70 - 0.80$\mu m$) each with
2.5-3.5 m resolution to produce a composite color image.}. The
imaging system has a coverage area at 350 km of 28.3 $× $ 448
km, obtained with oscillation of the satellite by $± 30^o$ in
the cross-track direction.    Onboard memory capacity is  769
Gbit. The RF communications for the payload data are in X-band at
8.2-8.4 GHz with  a downlink data rate of 75, 150 and 300 Mbit/s.
 \pam\ data    amounts to about 16 Gbyte/day,  sent to ground and processed in NTsOMZ station in Moscow,
 where also   the data analysis and quicklook procedures for \pam\ are performed.

\subsection{Scintillator / Time of Flight system}
The scintillator system\cite{tof} provides trigger for the
particles and time of flight information for incoming particles.
There are three scintillators layers, each composed by two
orthogonal   planes  divided in various bars (8 for S11, 6 for
S12, 2 for S21 and S12 and 3 for S32 and S33) for a total of 6
planes  and 48 phototubes (each bar is read by two phototubes). S1
and S3 bars  are 7 mm thick and  S2 bars are 5 mm. Interplanar
distance between  S1-S3 of  77.3 cm  results in  a TOF
determination of 250 ps precision for protons and 70 ps for C
nuclei (determined with beam tests in GSI), allowing separation of
electrons from antiprotons up to $\simeq 1$ GeV and albedo
rejection. The scintillator system is also capable of providing
charge information up to $Z=8$.
\subsection{Magnetic Spectrometer}
The permanent magnet \cite{track} is composed of 5 blocks, each
divided in 12 segments of  Nd-Fe-B alloy with a residual
magnetization of 1.3 T arranged to provide an almost uniform
magnetic field along the $y$ direction. The size of the cavity is
$13.1× 16.1× 44.5\: cm^3$, with a mean magnetic field of
0.43 T. Six layers of $300\mu \: m$ thick double-sided microstrip
silicon detectors are used to measure particle deflection with
$3.0± 0.1\: \mu m$ and $11.5± 0.6 \: \mu m$ precision in the
bending and non-bending views. Each layer is made  by three
ladders, each composed by two  $5.33× 7.00\: cm^2$ sensors
coupled to a VA1 front-end hybrid circuit. Maximum Detectable
Rigidity (MDR) was measured on CERN proton beam to be $\simeq 1\:
TV$.

\subsection{Silicon Tungsten Calorimeter}
Lepton/Hadron discrimination is performed by the Silicon Tungsten
sampling calorimeter \cite{calo} located on the bottom of \pam\ .
It is composed of 44 silicon layers  interleaved by 22  0.26 cm
thick Tungsten plates. Each silicon layer is composed arranging
$3× 3$ wafers, each of $80\times  80\times  .380\: mm^3$ and
segmented in 32 strips, for a total of 96 strips/plane. 22 planes
are used for the X view and 22  for  the Y view in order to
provide topological and energetic information of the shower
development in the calorimeter. Tungsten was chosen in order to
maximize electromagnetic radiation  lengths (16.3 $X_o$)
minimizing hadronic interaction length (0.6 $\lambda_{int} $). The
CR1.4P ASIC chip is used for front end electronics, providing a
dynamic range of 1400 mips (minimum ionizing particles) and
allowing nuclear identification up to Iron.
\subsection{Shower tail scintillator}
This  scintillator ($ 48\times 48\times 1\: cm^3$)   is located
below the calorimeter and is used to improve hadron/lepton
discrimination by measuring the energy not contained in the shower
of the calorimeter. It can also function as a standalone trigger
for the neutron detector.
\subsection{Neutron Detector}
The $60 \times  55 \times 15\: cm^3$ neutron  detector is composed
by 36 $^3He$ tubes arranged in two layers and surrounded by
polyethylene shielding and a 'U' shaped cadmium layer to remove
thermal neutrons not coming from the calorimeter. It is used to
improve lepton/hadron identification by detecting the number of
neutrons produced in the hadronic and electromagnetic cascades.
Since the former have a much higher neutron cross section than the
latter, where neutron production comes essentially through nuclear
photofission, it is estimated that \pam\ overall identification
capability is improved by a factor 10. As already mentioned, the
neutron detector is   used  to measure neutron field in Low Earth
Orbit and in case of solar particle events as well as in the high
energy lepton measurement.
\subsection{Anticoincidence System}
To reject  spurious triggers due to interaction with the main body
of the satellite, \pam\ is shielded by a number of scintillators
used with anticoincidence functions\cite{anti}\cite{pearce}. CARD
anticoincidence system is composed of four 8 mm thick
scintillators located in the area between S1 and S2. CAT
scintillator is placed on top of the magnet: it is composed by a
single piece with a central
 hole where the magnet cavity is located and  read out by 8 phototubes. Four scintillators,  arranged on the sides of the magnet, make the CAS
 lateral anticoincidence system.

\section{Integration, Launch and Commissioning}

\begin{figure}
\begin{center}
 \includegraphics[width=.55\textwidth, angle=-90]{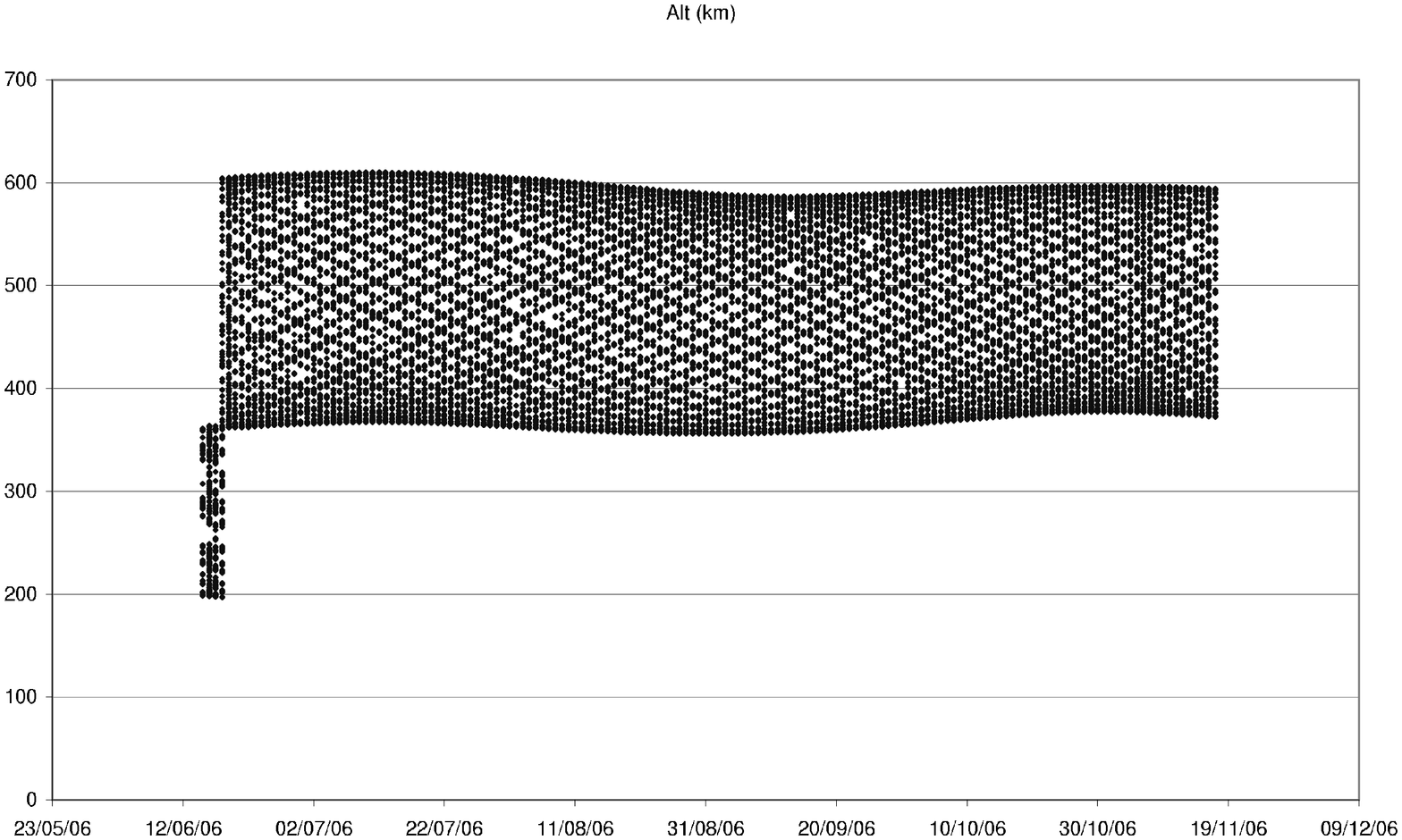}
\caption{  Height of Resurs as a function of time. After four days
in a parking orbit with $ 198 \times  360$ km the orbit was
boosted to $360  \times  604$ km. As of 17/11/2006 height has
passed to $372  \times  594 $ km.  } \label{height}
\end{center}
\end{figure}

\begin{figure}
\begin{center}
 \includegraphics[width=0.8\textwidth]{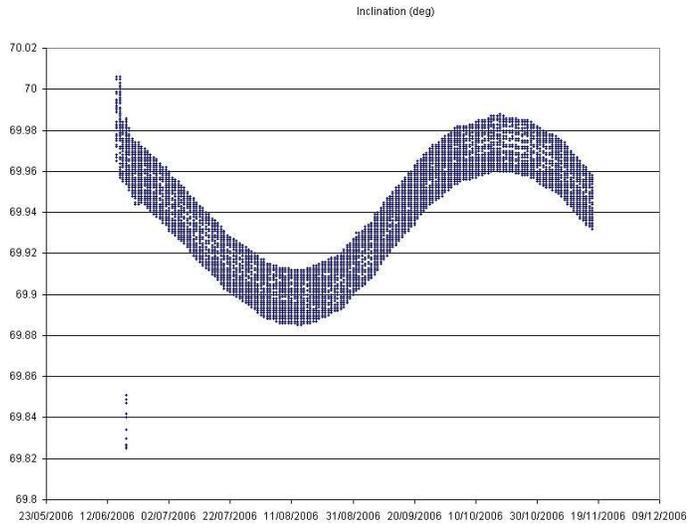} \caption{Inclination of
Resurs satellite as a function of time. The final boost after
launch increased inclination of the satellite. It is possible to
see secular oscillation of $\simeq 0.1^o$ and short term (daily)
variation of $0.03^o$.  } \label{inclination}
\end{center}
\end{figure}

\begin{figure}
\begin{center}
\includegraphics[width=0.8\textwidth]{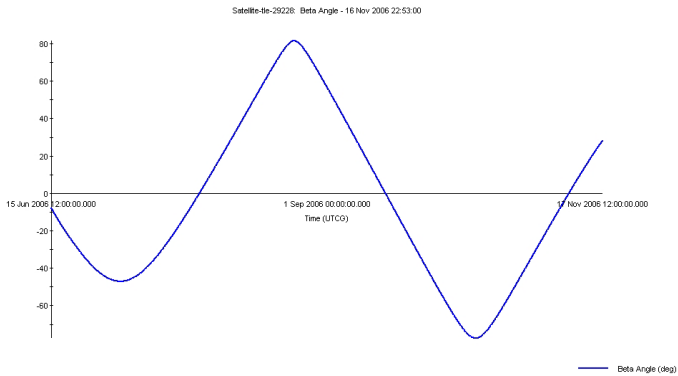}
\caption{Beta angle of satellite vs time. The inclined  orbit of
the satellite and the oblateness of the Earth result in the
precession of the node line resulting in a faster oscillation of
the angle.} \label{beta}
\end{center}
\end{figure}

Pamela was integrated in INFN - Rome Tor Vergata clean room
facilities; tests involved first each subsystem separately and
subsequently  the   whole apparatus simulating all interactions
with the satellite using an Electronic Ground Support Equipment.
Final tests involved cosmic ray acquisitions with muons for a
total of about 480 hours. The device was then   shipped to TsKB
Progress plant, in Samara (Russia), for installation in a
pressurized container  on board the Resurs-DK satellite for final
tests. Also in this case acquisitions with cosmic muons (140
hours) have been performed and have shown the correct functioning
of the apparatus, which was then integrated with the pressurized
 container and the satellite. The detector was then dismounted from the satellite and shipped by air to Baikonur
 cosmodrome (Kazakstan) where the the final integration phase took place in 2006.
\newline
The Soyuz-U rocket was launched  from Baikonur Cosmodrome  Pad 5
at Site 1, the same used for manned Soyuz and
 Progress cargoes to the International Space Station.
 Launch occurred    on June $15^{th}$ 2006,
 08:00:00.193 UTC with the payload reaching orbit after 8 minutes. Parking orbit had a semimajor axis
 of 6642 km. Final boost occurred on June $18^{th}$ 2006 when the orbit was raised with two engine firings to a semimajor axis of 6828 km. The maneuver was completed before 17:00 Moscow time.
  The  transfer orbit  resulted   in a height  increase from $198\times 360$ km   to $360 \times 604$ km, with the  apogee of the lower orbit becoming  perigee of the
final orbit. Also inclination of the satellite (Figure
\ref{inclination}) was increased from $69.93^o$ to $\simeq
69.96^o$. In the same Figure it is also possible to see long term
variations of $0.1^o$ in a period of 5 months due to the
oblateness of the Earth. In  Figure \ref{height} it is possible to
see the altitude of the satellite after launch, showing  the final
boost and the secular variation due to atmospheric drag, resulting
in a decrease of the apogee of 10 km in 5 months and a
corresponding increase of the number of revolutions/day
(spacecraft velocity is inversely proportional to square root of
height).
 To compensate for
atmospheric drag, the altitude of the satellite is   periodically
reboosted by vernier engines. To perform this maneuver  the
pressurized container housing \pam\ is folded back in the launch
position, the satellite is rotated $180^o$ on its longitudinal
axis and then engines are started. Reboost frequency depends from
orbital decay, due to atmospheric drag. Up to December  2006 the
activity has been low with two
  small Solar Particle Events in summer  and three larger events
   generated by sunspot 930 in December, so there has not been the need
to perform this maneuver so far. In Figure \ref{beta} is shown the
value  of the angle (Beta angle) between the orbital plane and the
Earth-Sun vector.  This value should vary with a one-year
periodicity but the oblateness of the Earth causes to precess with
a higher frequency. The position of the orbital plane  affects the
irradiation and temperature of the satellite, which is  - for
instance -  always under the Sun for high values of the absolute
value of beta.
 These thermal excursions are greatly reduced in the pressurized container of \pam\ thanks to the cooling loop
  with a fluid at a temperature of $28-33^o $
  which maintains the temperature of the detector relatively low  and reduces fluctuations within some
  degrees.

As already mentioned  Resurs-DK1 oscillates on its longitudinal
axis when performing Earth observations:   a detailed information
of the attitude of the satellite is provided to the CPU of \pam\
in order to know the orientation of the detector with precision of
$\simeq 1$ degree. Position and attitude information of the
satellite are provided to \pam\ CPU via a 1553 interface (used
also for Command and Control) and are based on the GLONASS (GLObal
Navigation Satellite System), similar to the GPS positioning
system. \newline On June 22, ground control successfully tested
the Geoton-1 optical-electronic system and the Sangur-1 data
receiving and processing system, according to Roskosmos. On June
23, 2006,  NTsOMZ   received first images from the satellite:  the
satellite conducted two photographic sessions, lasting five
seconds each. On September 15, 2006, Roskosmos announced that
testing of the spacecraft was successfully completed on that day
and State Commission planned to convene on September 21, 2006, to
declare the satellite operational. On September 22, 2006,
Roskosmos confirmed that the spacecraft was declared operational
as scheduled. Commissioning of the experiment proceeded in
parallel with Resurs-DK1 and mostly consisted in a fine tuning of
the observational parameters of \pam\  and the on board software,
optimizing time and schedule of downlinks to maximize live time of
the instrument.

\section{In flight data and instrument performance in Low Earth Orbit }

\begin{figure}[h!]
\begin{center}
  \includegraphics[width=\textwidth]{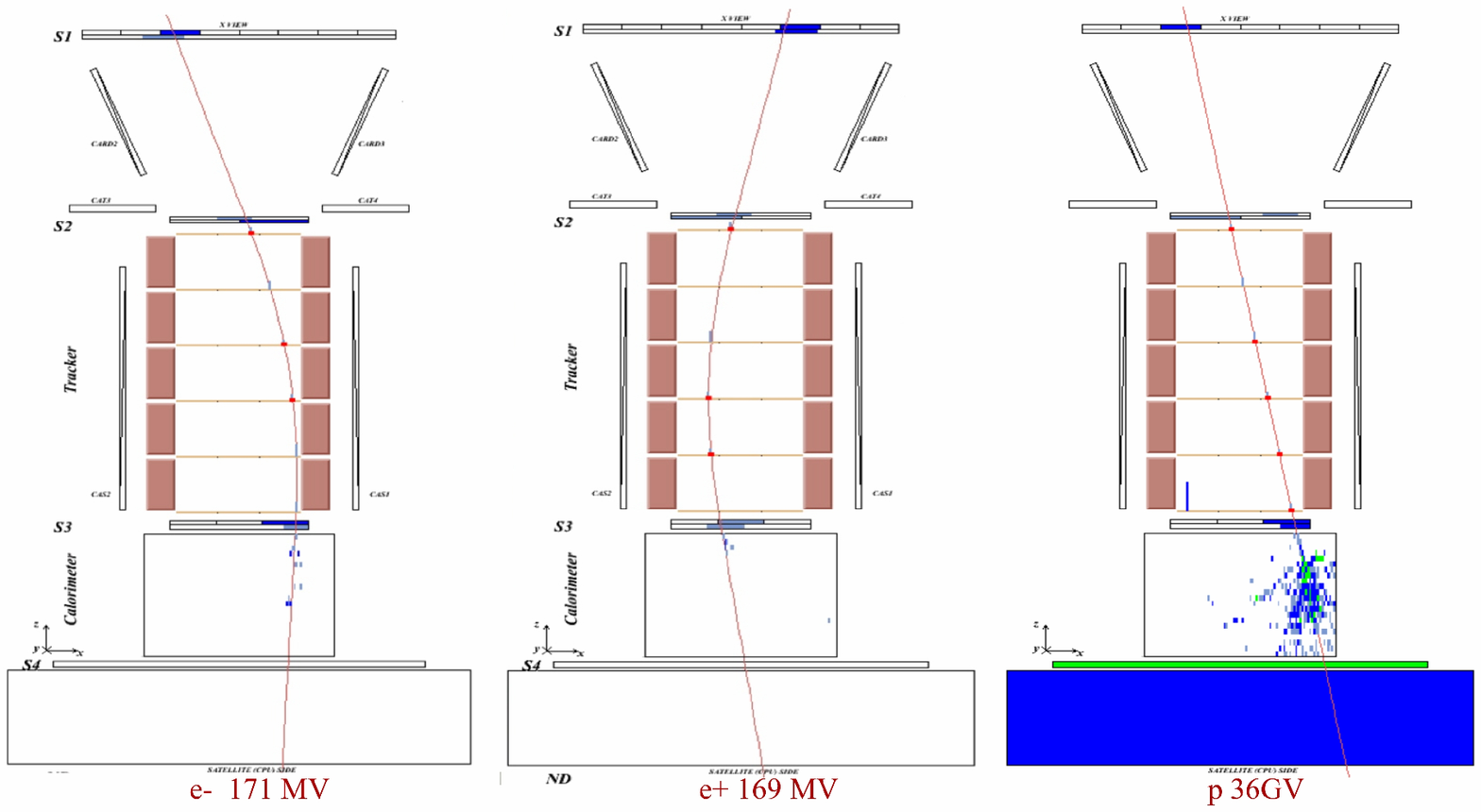}
\end{center}
\caption{Some cosmic ray events observed with \pam . Left:  0.171
GV $e^-$. The particle enters the detector from the top hitting
the two layers of S1 and the two layers of S2, located just above
the magnet cavity. The trajectory is bent by the magnetic field
and its rigidity is revealed by the microstrip detector of the
tracker. The particle interacts with the bottom scintillator (S3)
before absorption by the Si-W tracking calorimeter. Centre:  0.169
GV positron. Aside from the opposite curvature, the particle
interacts as in the preceding case. Right:  36 GV proton. Its high
rigidity reduces the magnet curvature. The  calorimeter shows the
shower from an hadronic interaction, with secondary particles
hitting the shower tail scintillator (S4) and the neutron
detector. } \label{eleeve}
\end{figure}

\pam\ was first switched on  June, $26^{th}$ 2006. Typical events are shown in Figure \ref{eleeve} where   an
electron and a positron crossing the detector and being bent in
different directions by the magnetic field are shown. In the third
panel a proton interacting hadronically in the calorimeter  is
visible. Note that the two leptons have  energies  too low to give
appreciable electromagnetic showers.

\begin{figure}[h!]
\begin{center}
  \includegraphics[width=\textwidth]{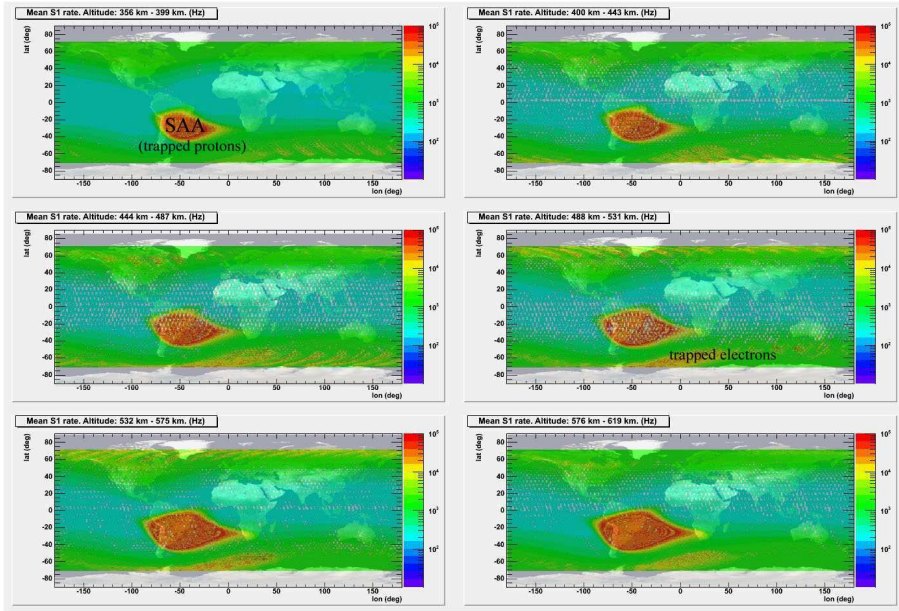}
\end{center}

\caption{All particle map ($E>35$ MeV  p; $E>3.5$ MeV  $e^-$)
measured at various altitudes with \pam . In it are visible the
proton (equatorial) and electron (high latitude) radiation belts,
regions of trapped particles where the flux can increase several
orders of magnitude. The size of the belts increases with altitude
where the weaker magnetic field is capable of trapping lower
energy particles. } \label{mappepamela}
\end{figure}

 In Figure
\ref{mappepamela} are  shown \pam\ world particle rate for S11*S12
at various altitudes (integral fluxes of $E>35$ MeV  p; $E>3.5$
MeV $e^-$), showing the high latitude electron radiation belts and
the proton belt in the South Atlantic Anomaly. Outside the SAA it
is possible to see the increase of particle rate at the
geomagnetic poles due to the lower geomagnetic cutoff. The highest
rates are found when the satellite crosses the trapped components
of the Van Allen Belts in agreement with AP-8 and AE-8 models for
trapped radiation\cite{ae8}.

\begin{figure}[h!]
\begin{center}
\includegraphics[width=.8\textwidth]{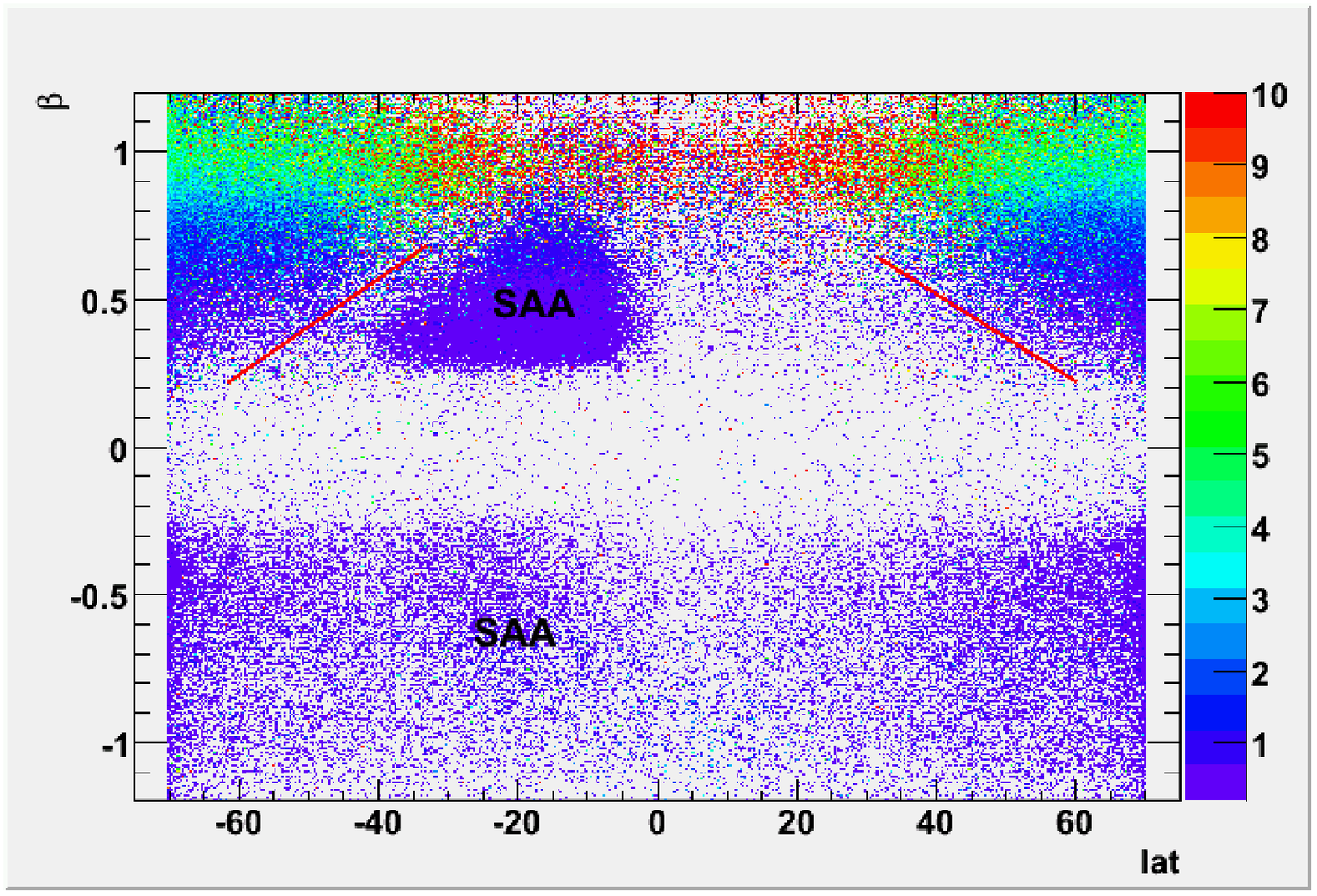}
\end{center}
 \caption{$\beta$ vs geographical latitude  of particles measured
 with \pam . Color code represents rigidity measured with the
 tracker. The red lines are to guide the eye and show the cutoff on
 galactic particles.  High  rigidity particles are present at all
 latitudes, whereas lower $\beta$ events (mostly due to protons)
 are observed only at high latitudes and in the SAA.}
\label{betarig}
\end{figure}

In Figure \ref{betarig} is shown the $\beta = v/c$ of particles
measured with the Time of Flight (TOF) system as function of the
geographical latitude observed. It is possible to see the effect
of geomagnetic  cutoff   on low energy particles, present only
closer to the poles. Also the South Atlantic region,  composed
mostly of  low energy ($E< 200 MeV$), low $\beta $  trapped
protons is clearly seen at the latitudes between 40$^o$ and 20$^o$
 S.
 Also   albedo ($\beta <0$)
particles crossing  the detector from the bottom to the top are
shown in the plot. Note   the absence of high energy albedo
particles.

\begin{figure}[h!]
\begin{center}
\hspace{.5cm}\includegraphics[width=.8\textwidth]{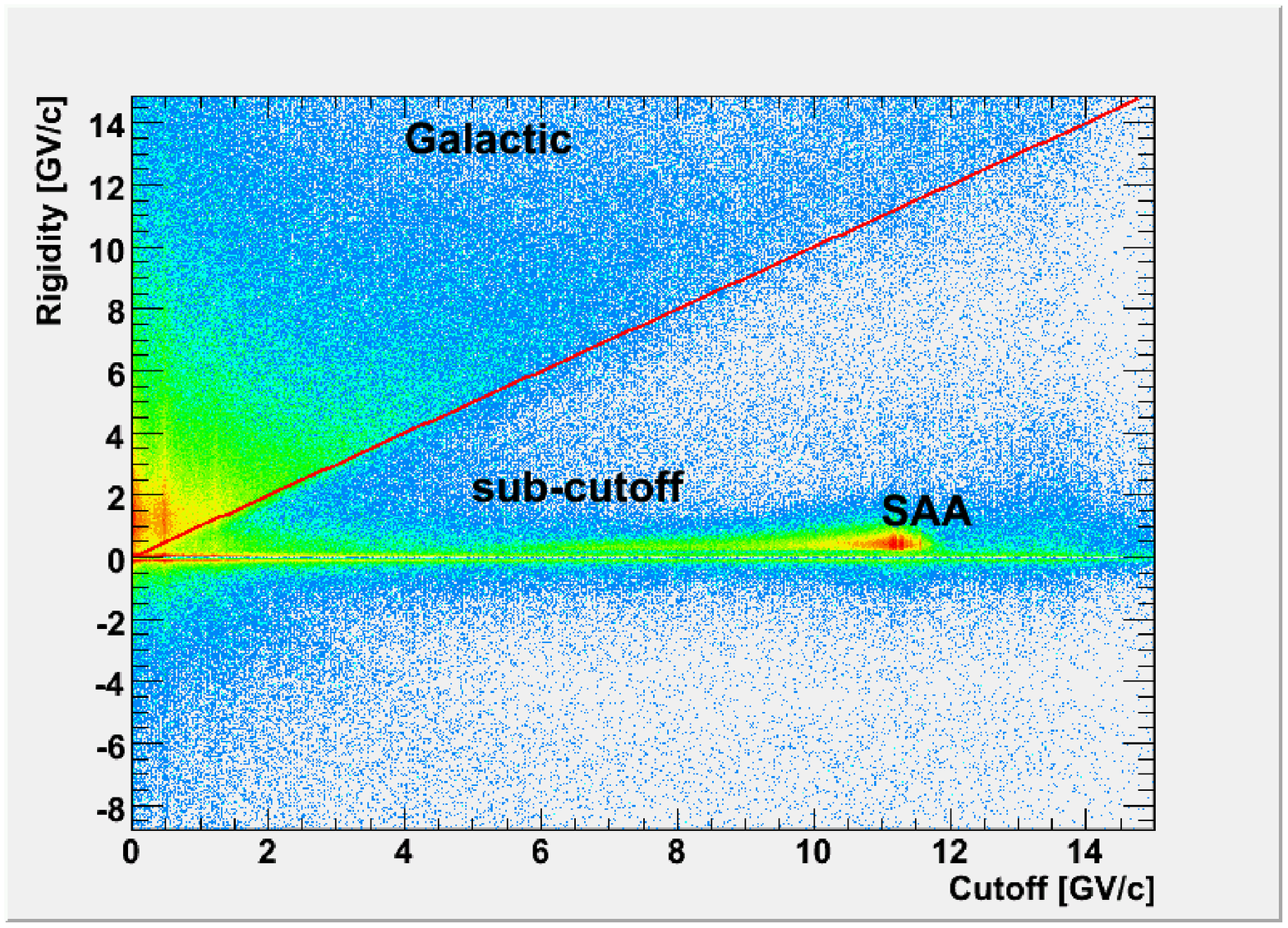}
\end{center}
\caption{Rigidity vs Stormer Cutoff observed with \pam . Colour
bar represents $\beta $ of particles measured from the TOF. The
effect of the geomagnetic field on galactic particles is clearly
visible. Primary particles have an energy above the cutoff and are
well separated from reentrant albedo events produced in the
interaction of particles with the Earth's
atmosphere.}\label{rigcutoff}
\end{figure}

\begin{figure}[!ht]
\begin{center}
\includegraphics[scale=.4]{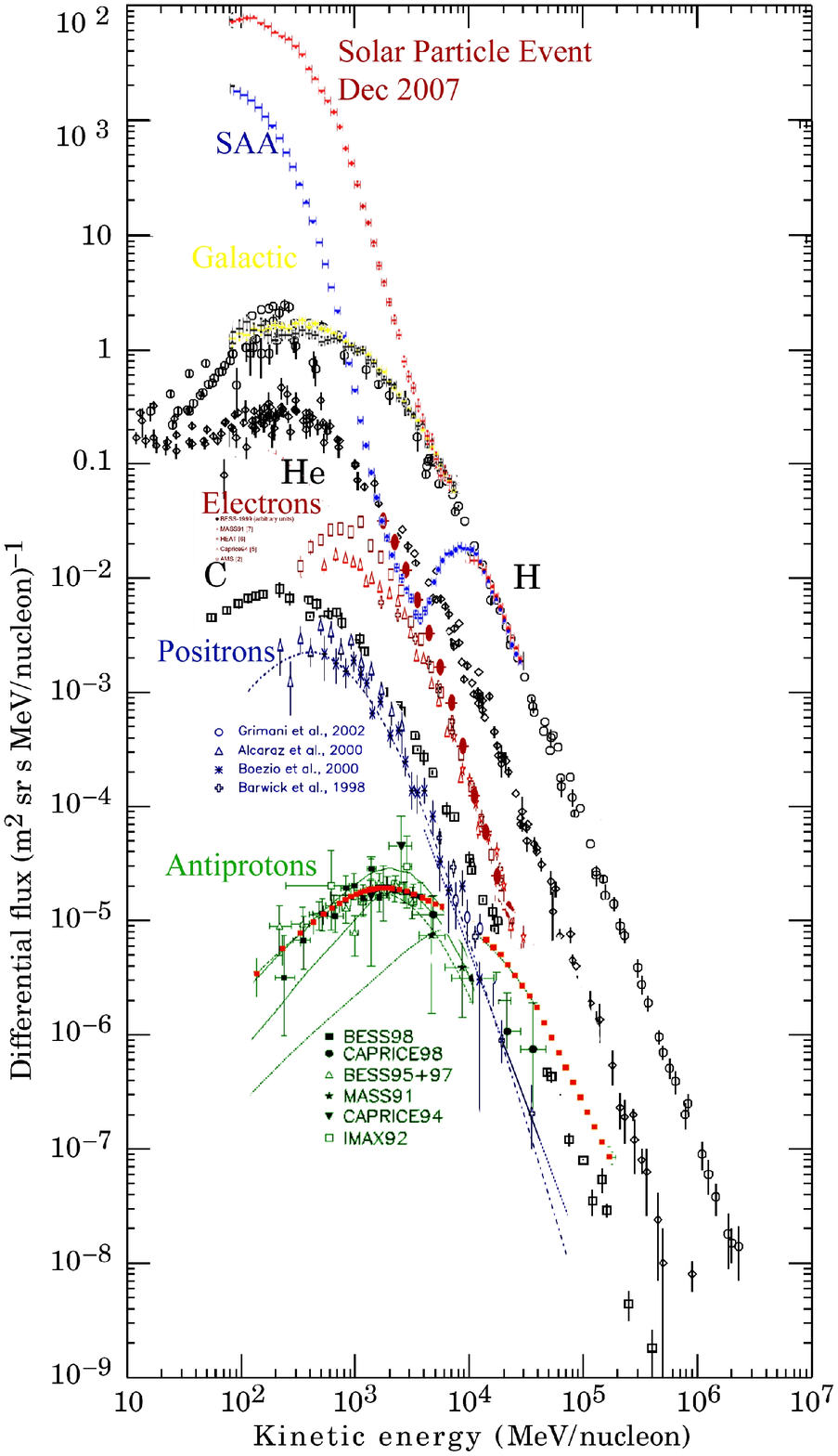}
\end{center}
\caption{Differential energy spectra of the different particles
detectable by \pam . Protons and Helium nuclei dominate the
positive  charge spectrum and electrons the negative charge
spectrum. Antiparticles are extremely rare in cosmic rays, with
positrons as abundant as Carbon nuclei. \pam acceptance energy
range is
     80 MeV -   190 GeV for antiprotons and   50 MeV - 270 GeV for positrons.
On the experimental data for antiproton spectra is shown an
expected contribution in case of a 964 GeV neutralino. Most
intense fluxes refer to the trapped protons in the South Atlantic
Anomaly and those coming from the December 13, 2006 Solar Particle
event. } \label{tuttiirc}
\end{figure}

\section{Scientific Objectives and first observations}

\pam\ can perform a detailed measurement of the composition and
energy spectra of cosmic rays of galactic, trapped and secondary
nature in Low Earth Orbit. Its $70^o$, $350\times 600 $ km orbit
makes it particularly suited to   study items of galactic,
heliospheric and trapped nature. Furthermore, the long duration of
the mission and the orbit configuration  should allow for studies
of spatial and temporal  dependence in solar quiet and active
conditions \cite{cosp08,cosp08jov,cosp08helio}. Indeed
for its versatility and detector redundancy \pam\ is capable to
address at the same time a number of different cosmic ray issues
 ranging over a very wide energy range, from the trapped particles in the Van Allen Belts, to electrons of Jovian origin, to the study of the
 antimatter component.
Figure \ref{tuttiirc} shows the different components of the cosmic
ray particle and antiparticle  fluxes with some of the \pam\
measurements.   Galactic protons are dominant, with Solar
Energetic and trapped particles being the only components more
abundant, albeit in an interval of time and in a specific region
of the orbit respectively.
 Here we briefly describe the main scientific objectives of the experiment and some of the preliminary  results obtained up to now.

\subsection{Antimatter research.} 
\begin{figure}[!ht]
\begin{center}
\includegraphics[width=.6\textwidth]{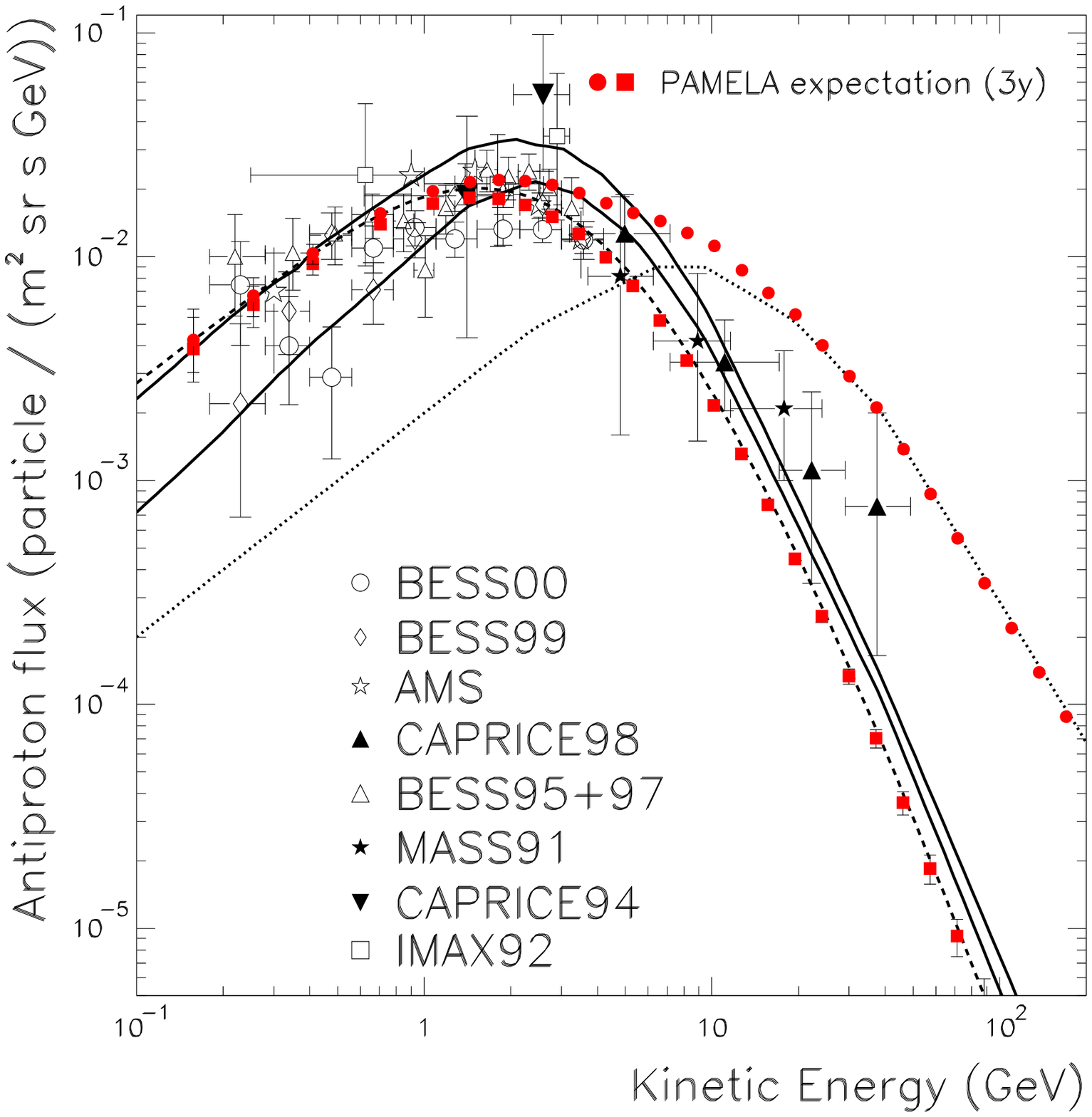}
  \caption{ Recent experimental \antip\
spectra (BES\-S00 and BES\-S99 ~\cite{asa02}, AMS~\cite{agu02},
CAPRICE98~\cite{boe01a}, BES\-S95+97~\cite{ori00},
MASS91~\cite{bas99}, CAPRICE94~\cite{boe97}, IMAX92~\cite{mit96})
along with theoretical calculations for pure \antip\ secondary
production (solid lines: \cite{sim98}, dashed line: \cite{ber99b})
and for pure \antip\ primary production (dotted line:
\cite{ull99}, assuming the annihilation of neutralinos of mass
964~GeV/c$^2$). (Taken from \cite{Pi07})} \label{pbflu}
\end{center}
\end{figure}

\begin{figure}[!ht]
\begin{center}
\includegraphics[width=.6\textwidth]{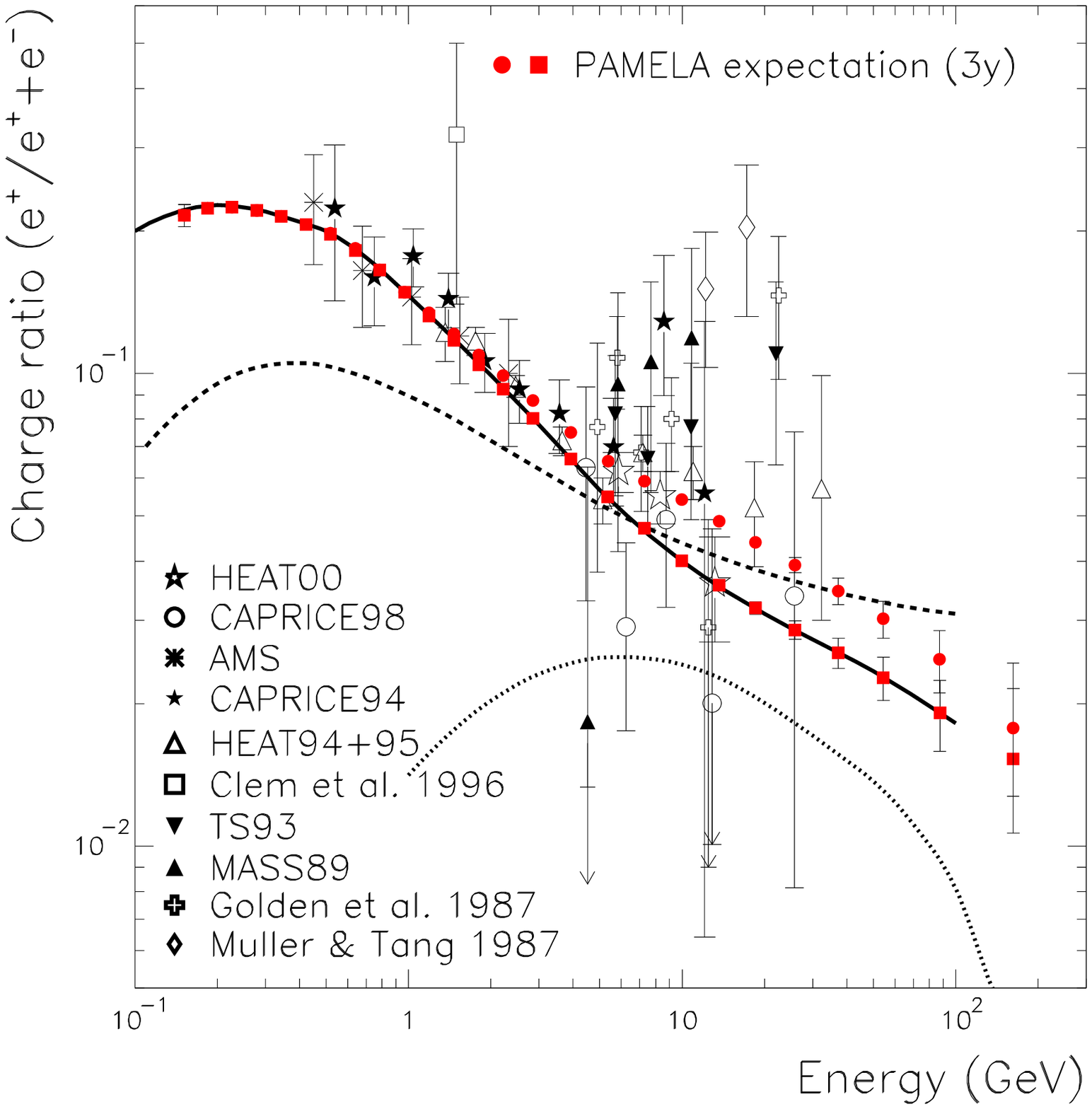}
  \caption{    The positron fraction as a function
of energy measured by several experiments
(\cite{gol87,mul87,cle96} and MASS89~\cite{gol94},
TS93~\cite{gol96}, HEAT94+95~\cite{bar98}, CAPRICE94~\cite{boe00},
AMS~\cite{alc00}, CAPRICE98~\cite{boe01b}, HEAT00~\cite{bea04}).
The dashed \cite{pro82} and the solid \cite{mos98} lines are
calculations of the secondary positron fraction. The dotted line
is a possible contribution from annihilation of neutralinos of
mass 336~GeV/c$^2$ \cite{bal99}. The expected PAMELA performance,
in case of a pure secondary component (full boxes) and of an
additional primary component (full circles), are indicated in both
panels. Only statistical errors are included in the expected
PAMELA data. Taken from \cite{Pi07}.} \label{pbflu2}
\end{center}
\end{figure}

The study of the antiparticle component (\antip, \posit)
  of cosmic rays  is the main scientific goal of \pam.  A long term and detailed study of the antiparticle spectrum over a very wide energy spectrum
 will allow to shed light over several questions of cosmic ray physics,  from particle production and propagation in the galaxy to
 charge dependent modulation in the heliosphere to dark matter detection.  In Figure \ref{pbflu} and \ref{pbflu2} are  shown the current status of
  the antiproton and positron measurements compared with \pam\ expected measurements in three years. In each case the two curves refer
   to a secondary only hypothesis with an additional contribution of a neutralino  annihilation. Also cosmological
    issues related to detection of a dark matter signature and search for antimatter (\pam\  will search for ${\overline{He}}$ with a
    sensitivity of $\approx 10^{-8}$) will therefore be addressed with this device.

\subsubsection{\bf Antiprotons} \pam\ detectable energy spectrum of \antip\ ranges from 80 MeV  to 190 GeV. Although the quality of \antip\ data has been improving in the recent years,  a   measurement of the energy spectrum
 of \antip\ will allow to greatly reduce the systematic error between the different balloon measurements, to study the
  phenomenon of charge dependent solar modulation,  and will for the first time explore the energy range beyond $\simeq 4 0$ GeV.
Possible excesses over the  expected secondary spectrum could be
attributed to neutralino annihilation; \cite{profumo,Ullio,donato}
  show that \pam\ is capable of detecting an excess of antiprotons due to neutralino annihilation in models  compatible with
   the WMAP measurements. Also \cite{lionetto} estimate that \pam\ will be able to detect a supersymmetric signal in many minimal supergravity  (mSUGRA)  models.
   The possibility to extract a neutralino annihilation signal from the background depends on the parameters used, the   boost
   factor  (BF) and the galactic proton spectrum. Other scenarios
   \cite{khl1} \cite{khl2} suppose the existence of heavy
   neutrinos  or stable heavy particles as DM constituents. In
   \cite{cfs08} the preliminary results of \pam\ on \antip  are
   compared with other measurements to explore the possibility of
   DM signature in fermion 3-plet ot 5-plet scenarios and conclude
   the possibility to extract a signal in case of BF=10.
\\  Charge dependent solar modulation, observed with the BESS balloon flights at Sun field reversal
 \cite{asa02} and more recently on a long duration balloon flight \cite{besspolar04}
will be monitored
 during the period of recovery going from the $23^{rd}$ solar minimum going to the $24^{th}$ solar maximum.
Also the existence, intensity and stability of secondary
antiproton belts \cite{Mi03}, produced by the interaction of
 cosmic rays with the atmosphere will be measured.

\subsubsection{\bf Positrons}

A precise measurement of the positron energy spectrum is needed to
distinguish dark matter annihilation from other galactic sources
such as hadronic
 production in giant molecular clouds, $e^+/e^-$ production in nearby pulsars
  or decay from radioactive nuclei produced in supernova explosions.
An interesting feature of $e^+$ is that - as electrons - they lose
most of   length scales of a few kiloparsecs   (<50). The cosmic
positron spectrum is  therefore a  samples of only the local dark
matter distribution\cite{hb08}.
   \pam\ is capable to detect   \posit\ in the energy range   50 MeV to 270 GeV.
Possibilities for dark matter detection in the positron channel
depend strongly on the nature of dark matter, its cross section
and the local inhomogeneity of the distribution. Hooper and Silk
\cite{hooper05} perform different estimation of  \pam\ sensitivity
according to different hypothesis of the dark matter component:
detection is possible  in case of an  higgsino of mass up to 220
GeV (with BF=1) and   to 380 GeV (with BF=5). Kaluza Klein models
\cite{hooper2} would give a  positron flux above secondary
production  increasing above 20 GeV and thus clearly compatible
with \pam\ observational parameters. In case of a bino-like
particle, as supposed by Minimal Supersymmetric Standard Model,
\pam\ is sensible to cross sections of the order of $2-3\times
10^{-26}$ (again, depending of BF). In case of Kaluza Klein
excitations of the Standard Model the sensitivity of \pam\ is for
particles up to 350 and 550 GeV. In the hypothesis of the littlest
Higgs model with T parity, the dark matter candidate is a heavy
photon which annihilates mainly into weak gauge bosons in turn
producing positrons. In \cite{asano} is  shown that \pam\ will be
able to identify this signal if the mass of the particle is below
120 GeV and the BF is 5.  Hisano et al.,  \cite{hisano2006} assume
a heavy wino-like dark matter component, detectable with \pam\ in
the positron spectrum (and with much more difficulty in the
antiproton channel) for mass of the wino above 300 GeV. This model
predicts that if the neutralino has a mass of 2 TeV the positron
flux increases by several orders of magnitude due to resonance of
the annihilation cross section in $W^+W^-$ and $ZZ$: in this
scenario not only such a signal would be visible by \pam\ but also
be consistent with the increase of positrons measured by HEAT
\cite{heat}.
 In conclusion a detailed measurement of the
 positron spectrum, its spectral features and its dependence from solar modulation will either provide evidence for a dark matter signature or
 strongly constrain and discard many existing models.

\subsection{\bf Galactic Cosmic Rays}
Proton and electron spectra will be measured  in detail with \pam
. Also light nuclei (up to O) are   detectable with the
scintillator system. In this way it is possible to study with high
statistics the secondary/primary cosmic ray nuclear and isotopic
abundances such as B/C,  Be/C,   Li/C and $^3He/ ^4He$. These
measurements will constrain existing production and propagation
models in the galaxy, providing detailed information on the
galactic structure and the various mechanisms involved.

\subsection{\bf Solar modulation of GCR}

\begin{figure}[h!]
\begin{center}
\includegraphics[width=\textwidth]{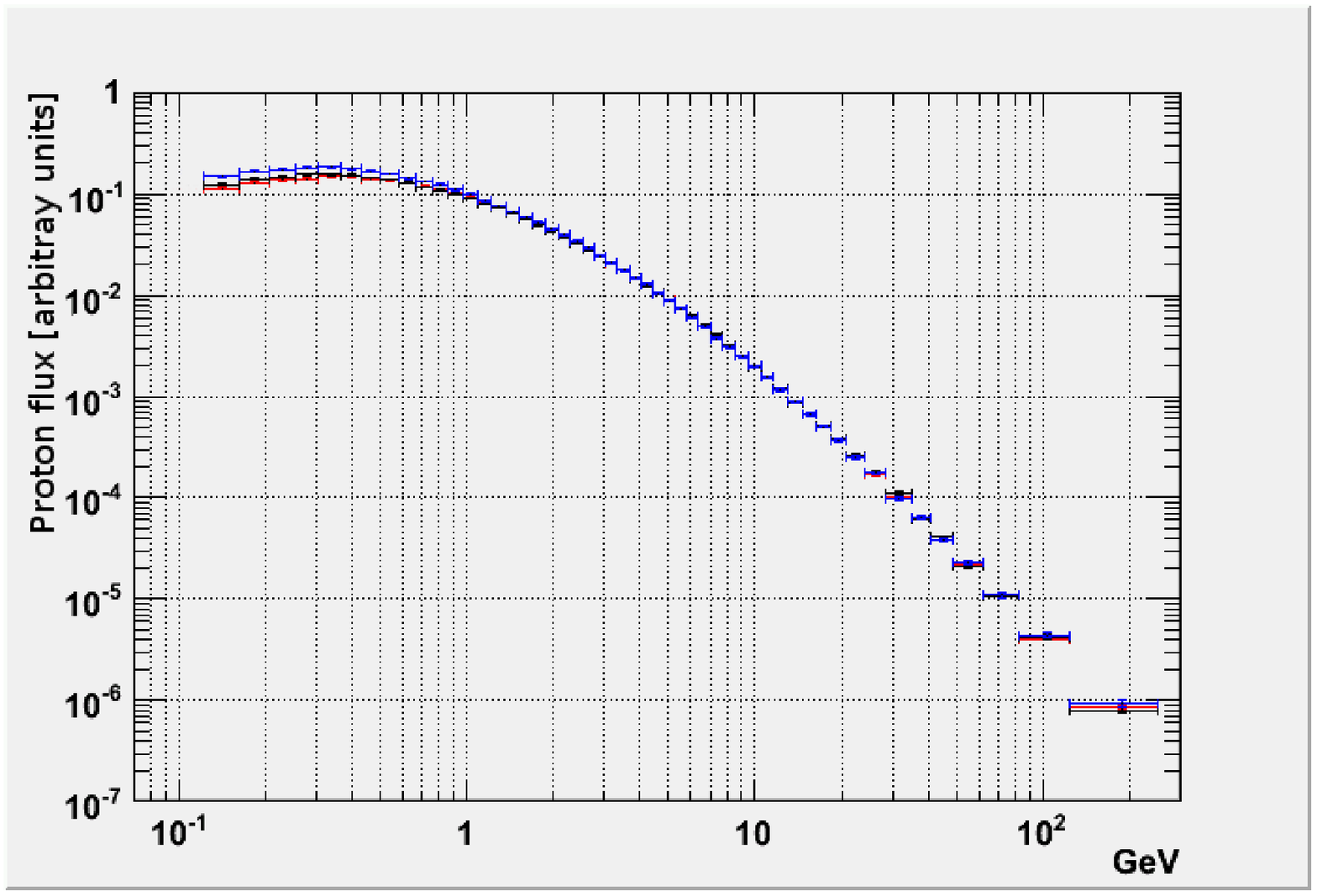}
\end{center}
 \caption{Differential spectrum of protons measured in July 2006 (red),
  January 2007 (black), August 2008 (blue). Below 1 GeV it is possible to see the         flux variation due to solar modulation.}
\label{solmodulation}
\end{figure}

Launch of \pam\ occurred   in the recovery  phase of solar minimum
with negative polarity (qA<0)  toward   solar maximum of cycle 24.
We are currently in an  unusually long solar minimum with
disagreement over prediction on the behavior of the intensity and
peaking time of next maximum.  In this period \pam\ has been
observing   solar modulation of galactic cosmic rays during
decreasing   solar activity. A long term measurement of the
proton, electron and nuclear flux at 1 AU can   provide
information on propagation phenomena occurring in the heliosphere.
As already mentioned, the possibility to identify the antiparticle
spectra will allow to study also charge dependent solar modulation
effects.  In Figure \ref{solmodulation} are shown the proton
fluxes measured in various periods of the solar minimum. It is
possible to see how the effect of decreasing solar activity on the
flux of cosmic rays is visible even during this solar quiet
period,   in agreement with the increase of neutron monitor
fluxes. Future work will involve correlation of the particle flux
and solar modulation with variation with time of tilt angle.

\subsection{\bf Trapped particles in the Van Allen Belts }

\begin{figure}[!ht]
\vspace{-.5cm}
\begin{center}
\includegraphics[width=.6\textwidth]{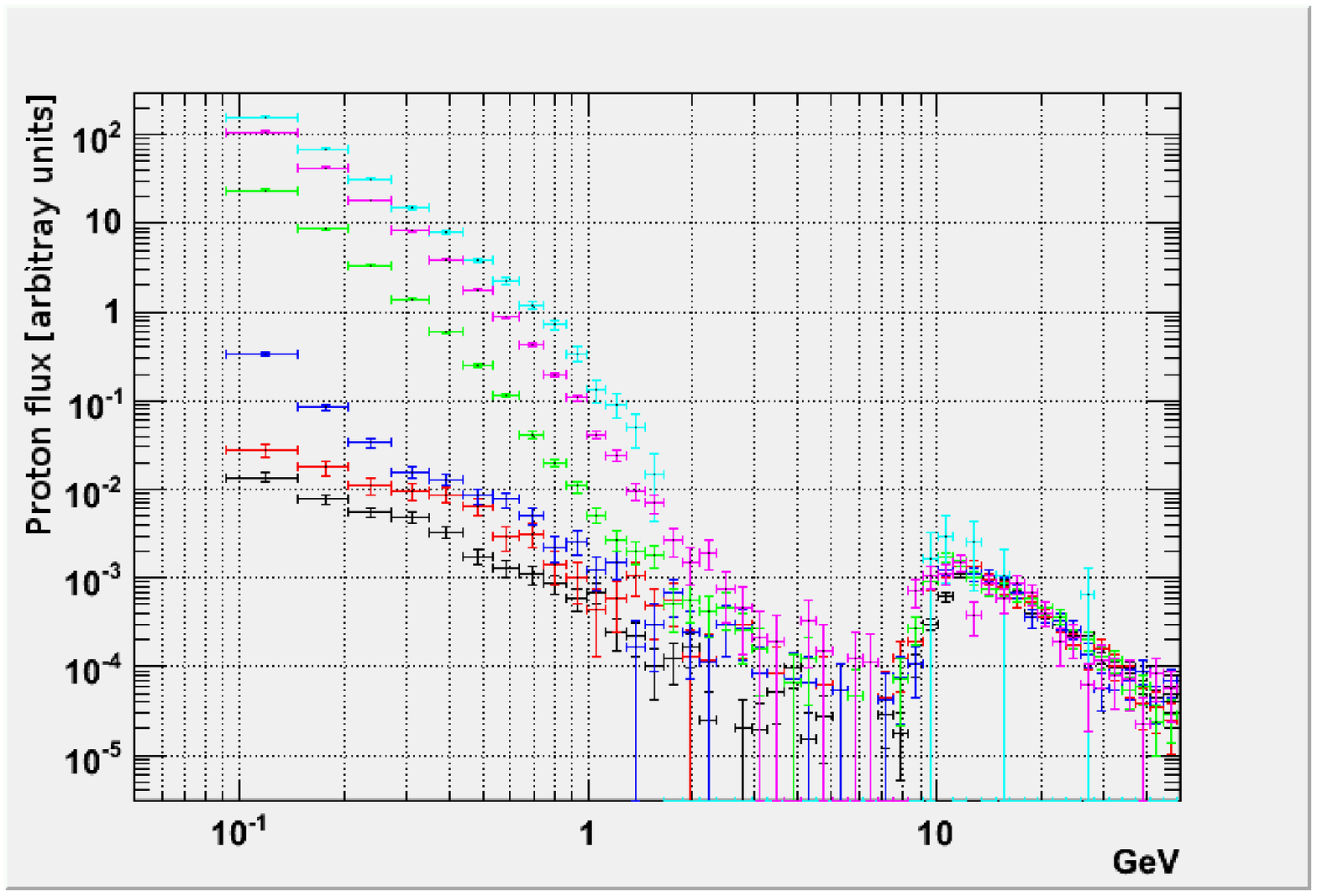}
\includegraphics[width=.6\textwidth]{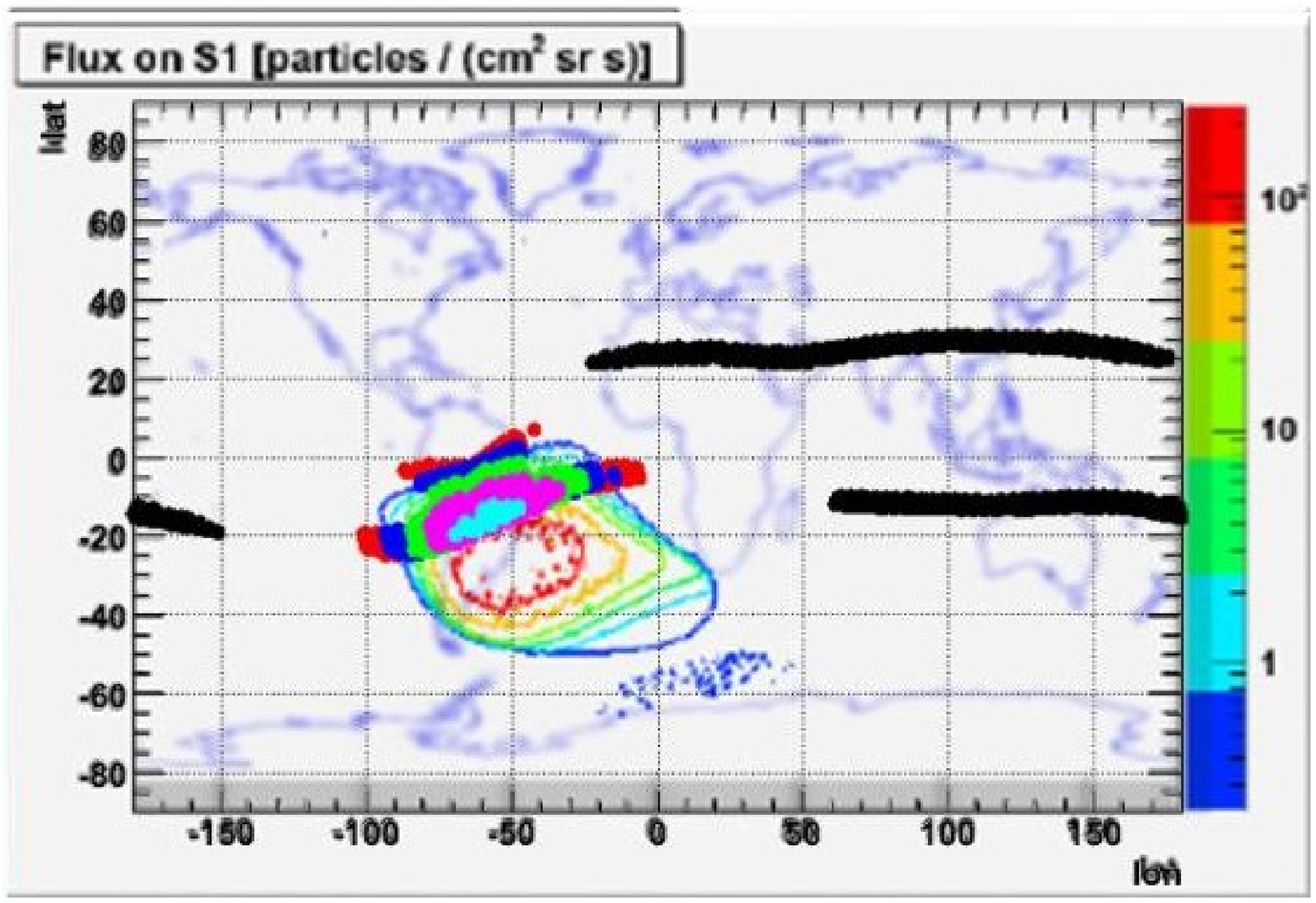}
\end{center}
\caption{Top: Plot of the differential energy spectrum of \pam\ in
different regions of the South Atlantic Anomaly. Regions are
selected according  to different intensity of the    magnetic
field (Black $B
> 0.3 G$ - outside the SAA, Red      $  0.22 G < B < 0.23 G$ Blue      $ 0.21 G < B
< 0.22 G $ Green       $ 0.20 G < B < 0.21 G $ Pink     $  0.19 G
< B < 0.20 G $ Turquoise    $0.19 G > B$ )  in the  cutoff region
$10.8 GV < Cutoff < 11.5 GV$. Trapped particles over the secondary
particle flux measured in the same cutoff region outside the
anomaly (black curve) are evident up to and above 1 GeV. Bottom:
geographical regions corresponding to the above selection. The
color bar corresponds to counting rate of the S1 (topmost)
scintillator.  Note the geographical shift between the peak of the
SAA spectrum at high energy and the peak of the scintillator
counting rate. } \label{southatlantic}
\end{figure}

The   $70^o$ orbit of the Resurs-DK1 satellite allows for
continuous monitoring of the electron and proton belts. The high
energy ($>80 MeV$) component of the proton belt, crossed in the
South Atlantic region will be monitored in detail with the
magnetic spectrometer. Using the scintillator counting  rates   it
will be possible to   extend measurements of the particle spectra
to lower energies using the range method. Montecarlo simulations
have shown that the coincidence of the two layers of the topmost
scintillator (S1) allows \pam\ to detect $e^-$  from 3.5 MeV and
$p$ from 36 MeV. Coincidence between S1 and the central
scintillator (S2) allows us to measure   integral spectra of  9.5
$e^-$ and 63 MeV p. In this way it will be possible to  perform a
detailed mapping of the Van Allen Belts showing spectral and
geometrical  features. Also the neutron component will be
measured, although some care needs to be taken to estimate the
background coming from proton interaction with the main body of
the satellite. In Figure \ref{southatlantic} is shown the
differential energy spectrum measured in different regions of the
South Atlantic Anomaly. It is possible to see   flux increase
toward the centre of the anomaly. Particle flux exceeds several
orders of magnitude the flux of secondary (reentrant albedo)
particles measured in the same cutoff region outside the anomaly
and is maximum where the magnetic field is lowest. However  this
is not the  location of the flux at lowest energies   according to
scintillator counting rate.  The reason for this difference is
currently under investigation with comparison with existing
models\cite{ae8}\cite{mewaldtprotons}\cite{mewaldtantiprotons}.

\subsection{Secondary particles production in the Earth's
atmosphere}

\begin{figure}[ht]
\vspace{-.5cm}
\begin{center}
\includegraphics[width=.8\textwidth]{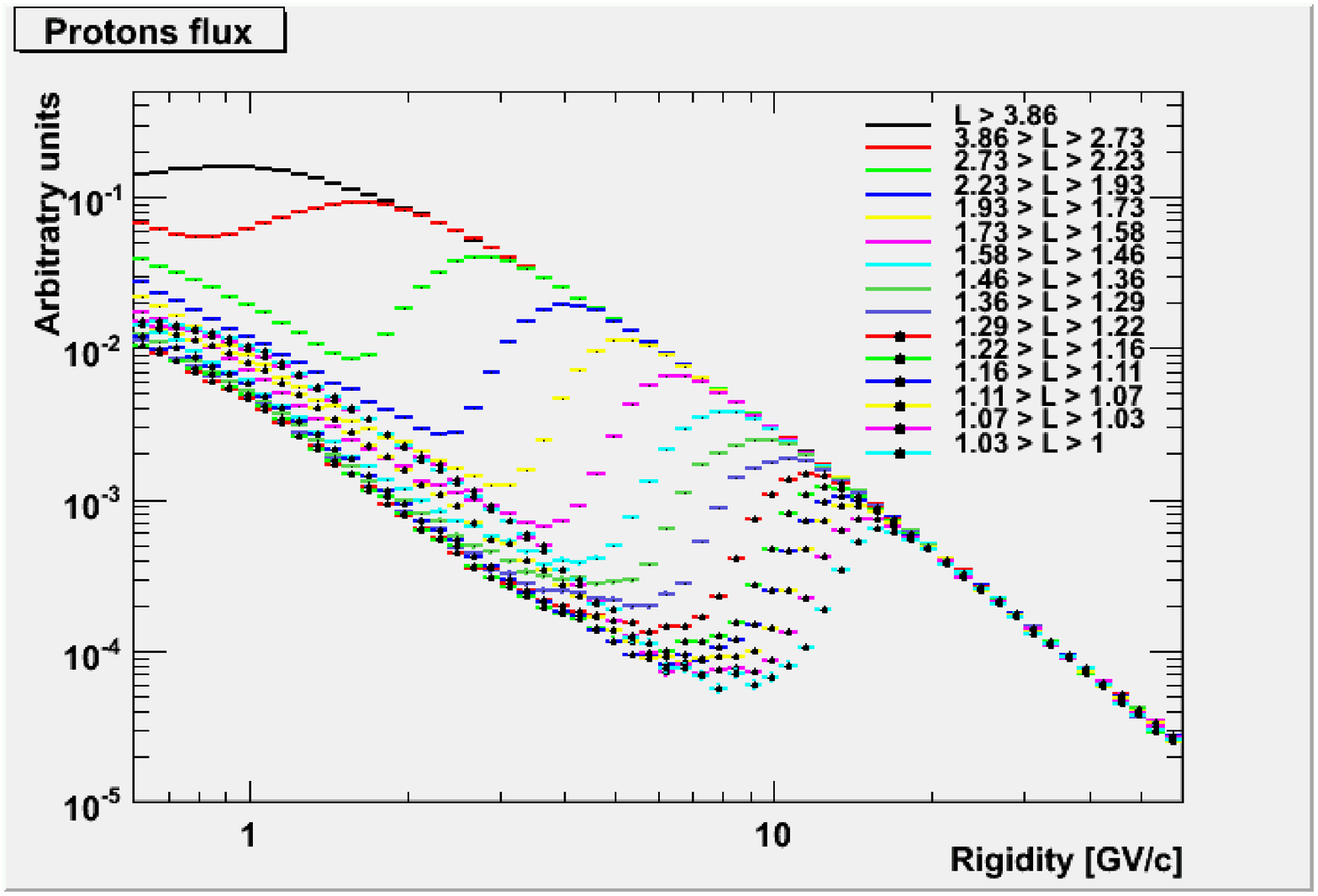}
\end{center}
\caption{Plot of the differential energy spectrum of \pam\ at
different L shells (according to McIlwain parameter). It is
possible to see the primary spectrum at high rigidities and the
reentrant albedo (secondary) flux at low rigidities. The
transition between primary and secondary spectra is lower at lower
cutoffs. }\label{subcutoff}
\end{figure}

To clearly separate primary component from the reentrant albedo
(particles produced in interactions of cosmic rays with the
atmosphere below the cutoff and propagating on Earth's magnetic
field line) component it is necessary to evaluate the local
geomagnetic cutoff. This is estimated using IGRF magnetic field
model along the orbit; from this the McIlwain $L$ shell is
calculated\cite{igrf}. In this work we have used the vertical
Stormer (defined as $G=14.9/L^2$) approximation\cite{shea}. Figure
\ref{rigcutoff} shows the rigidity of particles as function of the
evaluated cutoff $G$. The primary (galactic) component, with
rigidities above the cutoff  is clearly separated  from the
reentrant albedo (below cutoff) component, containing also trapped
protons in the SAA. Note that color code shows the absolute value
of $\beta $ so that negative rigidity particles in the SAA region
are albedo ($\beta <0$ protons) with negative curvature in the
tracker due to the opposite velocity vector. In Figure
\ref{subcutoff} is shown the particle flux measured at different
cutoff regions. It is possible to see the primary (above cutoff)
and the secondary (reentrant albedo - below cutoff ) component. At
the poles, where cutoff is below the detection threshold of \pam\
the secondary component is not present. Moving toward lower
latitude regions the cutoff increases and it is possible to see
the two components, with the position of the gap increasing with
the increase of the cutoff. An accurate measurement of the
secondary component is of   relevance both in the calculation of
the atmospheric neutrino\cite{honda2004}\cite{honda2007} flux and
in the estimation of hadronic cross sections (protons on O or N)
at high energies, not otherwise determinable on ground.

\subsection{\bf Solar energetic particles}

\pam\ observations are taking place at solar minimum, where about
10 significant solar events are expected during the three years
experiment's lifetime\cite{shea}. The observation of solar
energetic particle (SEP) events with a magnetic spectrometer will
allow several aspects of solar and heliospheric cosmic ray physics
to be addressed for the first time.

\subsubsection{Electrons and Positrons}
\label{sec:positron} Positrons are produced mainly in the decay of
$\pi^{+}$ coming from nuclear reactions occurring at the flare
site. Up to now, they have only been measured indirectly by remote
sensing of the gamma ray annihilation line at 511~keV. Using the
magnetic spectrometer of \pam\ it will be possible to separately
analyze the high energy tail of the electron and positron spectra
at 1 Astronomical Unit (AU) obtaining information both on particle
production and charge dependent propagation in the heliosphere in
perturbed conditions of  Solar Particle Events.

\subsubsection{Protons}
\label{sec:proton} \pam\ is capable to measure the spectrum of
cosmic-ray protons from 80~MeV up to almost 1~TeV and  therefore
will be able to measure the solar component over  a very wide
energy range (where the upper limit will be limited  by size and
spectral shape of the event). These measurements will be
correlated with other instruments placed in different points of
the Earth's magnetosphere to give information on the acceleration
and propagation mechanisms of SEP events. Up to now there has been
no direct measurement~\cite{miroshnichenko} of the high energy
($>$1~GeV) proton component of SEPs. The importance of a direct
measurement of this spectrum is related to the fact~\cite{ryan}
that there are many solar events where the energy of protons is
above the highest ($\simeq$100 MeV) detectable energy range of
current spacecrafts,  but is below the detection threshold of
ground Neutron Monitors~\cite{bazilevskaya}. However, over the
\pam\ energy range, it will be possible to examine the turnover of
the spectrum, where we find the limit of acceleration processes at
the Sun.

\subsubsection{Nuclei}
\label{sec:nuclear}  \pam\  can identify light nuclei up to Carbon
and isotopes of Hydrogen and Helium. Thus we can investigate the
light nuclear component related to SEP events over a wide energy
range. This should contribute to  establish whether there are
differences in the  composition of the high energy (1 GeV) ions
  to the low energy component ($\simeq$ 20 MeV)   producing $\gamma $ rays or the quiescent solar corona\cite{ryan05}.
 These
measurements will help us to better understand the selective
acceleration processes in the higher energy
impulsive~\cite{reames} events.

\subsubsection{Lowering of the geomagnetic cutoff}
\label{sec:geo} The high inclination of the orbit of the
Resurs-DK1 satellite will allow \pam\ to study
\cite{ogliore}\cite{leske} the variations of cosmic ray
geomagnetic cutoff due to the interaction of the SEP events with
the geomagnetic field.

\subsubsection{13 December 2006 Solar Particle event}

At the time of writing the most significant events detected by
\pam\ occurred between December 6$^{th}$ and 17$^{th}$  2006 and
were originated from region 930. Dec 6$^{th}$ event was originated
in the  East, resulting in a gradual proton event  reaching Earth
on Dec 7$^{th}$ and lasting until the   events  of Dec 13 and
14\cite{goes}.  On   13 December 2006, 02:38 UT  an X3.4/4B solar
flare occurred   in active region NOAA 10930 ($S06^oW23^o$).
   The interaction   between the fast
rotating sunspot and the ephemeral regions triggers continual
brightening and finally produces the major
flare\cite{zhangsongsep2006}. The intensity of the event (the
second largest GLE of cycle 23) is quite unusual for a solar
minimum condition. Starting at 2:50 UT on December 13, 2006,
various neutron monitors, with cutoff rigidities below about
$4.5\,GV$, recorded a Ground Level Enhancement (GLE70) with
relative increases ranging from $20\%$ up to more than $80\%$
(Apaty, Oulu) \cite{Bi07}\cite{Ta07}.  Apaty and Oulu also
registered the peak of the event beetween 02:40 UT and 03:10 UT,
while most of the neutron monitors had it between 03:10 UT and
03:40 UT.   The spectrum and its dynamic was investigated at
higher energies using ground measurements by neutron monitors at
different cutoff rigidities \cite{Va07} resulting in a spectral
estimation of $\gamma = 6$. The onset time was later for the
proton channels on-board of GOES-11 satellite: 03:00 UT for
greater than 100 MeV protons and 03:10 for greater than 10 MeV
protons \cite{Ta07}. \pam was in an high cutoff region at the
flare occurrence and reached the South Polar region   at about
03:10 UT. Muon monitors were also able to detect the GLE event and
its spatial-angular anisotropy has been measured \cite{Ti07}.
Differential proton spectra were directly meausured by GOES, ACE,
Stereo, SAMPEX at energies below $400\, MeV$.  With these
instruments it was also possibile to measure the elemental
composition of the various events\cite{Me07}\cite{Co07}.
\par
\begin{figure}[!ht]
\begin{center}
\includegraphics[width=.8\textwidth]{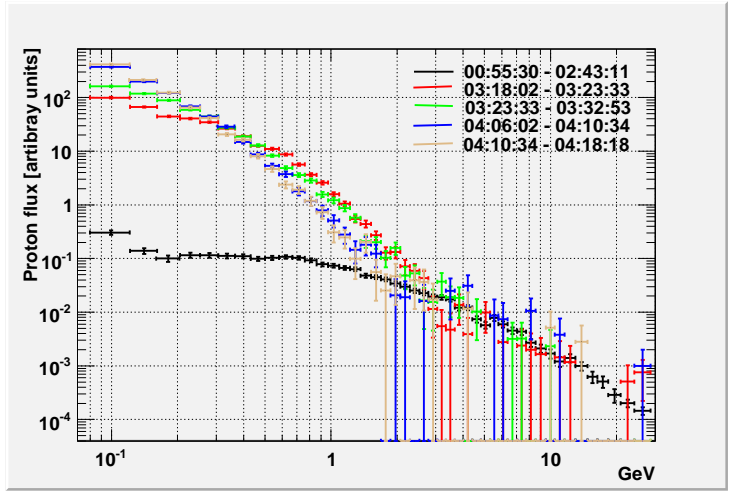}
\end{center}
\caption{Proton differential energy spectra in different time
intervals during the event of the 13th December 2006.  The black
line is the spectrum before the arrival of the charged particles
with a small peak at low energy due to the presence of solar
protons from previous events.  It can be observed that the maximum
flux of the high energy component of the solar protons arrives at
the beginning of the event while only one hour later the maximum
flux at low energy is detected.  On the other hand, the flux at
high energy decreases faster than at low energy.} \label{13eve}
\end{figure}

  The
event produced also a full-halo Coronal Mass Ejection (CME) with a
projected speed in the sky of 1774 km/s \cite{mmcme}. The forward
shock of the CME reached Earth at 14:38 UT on December 14, causing
a Forbush decrease of galactic cosmic rays which lasted for
several days. A second SPE of lower intensity and energy occurred
in conjunction with a X1.5 flare from the same active region (NOAA
10930, $S06^oW46^o$). A fourth event was observed at 17:23 UT on
December 16 by ACE with the downstream passage of the CME. In
Figure \ref{13eve} is shown the differential energy spectrum
measured with \pam\ in different periods of the event of the 13
December. It is possible to see that the event produced
accelerated  particles up to 3-4 GeV. A second smaller event
occurred on Dec 14, superimposing on the Forbush decrease caused
by the Coronal Mass Ejection  of the previous event reaching
Earth. Galactic particle flux thus decreased   in the energy range
up to 3 GeV, whereas solar particles were accelerated up to 1 GeV
for this event. The decrease was also observed by Wind, Stereo and
Polar but not by the GOES satellites, with the exception of   some
variation in the 15-40 MeV channell of GOES-12 \cite{Mu07}.
  The relative decrease record by \pam was up to $20\%$, depending on the energy.

\subsection{\bf High energy lepton component}
The calorimeter can provide an independent trigger to \pam\ for
high energy releases due to showers occurring in it: a signal  is
generated  with the release of energy above 150 mip in  all the 24
views of planes from 7 to 18. With this requirement the
geometrical factor of the calorimeter self-trigger is 400 $cm^2
sr$  if events  coming from the satellite are rejected. In this
way it is possible to study the    electron and positron flux in
the energy range between 300 GeV  and 2 TeV, where measurements
are currently scarce \cite{koba}. At this energy discrimination
with hadrons is performed with topological and energetic
discrimination of the shower development in the calorimeter
coupled with neutron information coming from the neutron detector.
This is because   neutron  production cross-section  in an e.m.
cascade is lower than in a hadronic cascade\cite{galper}.

\section{Conclusions}
\pam\ was successfully launched on June 2006 and is  currently
operational in Low Earth Orbit. The satellite and the detectors
are   functioning correctly. It it expected that data from \pam\
will provide information on several items of cosmic ray physics,
from antimatter to solar and trapped particles.



%
\end{document}